\documentclass[10pt,twoside]{article}
\usepackage{newpasp}
\usepackage{epsfig}

\newcommand{\Msun}{\ensuremath{M_\odot}}

\begin{document}
\title{Diffusive Nuclear Burning in Neutron Star Envelopes}
 \author{Philip Chang}
\affil{Department of Physics, Broida Hall, University of California,
Santa Barbara, CA 93106; pchang@physics.ucsb.edu}
\author{Lars Bildsten}
\affil{Kavli Institute for Theoretical Physics and Department of
Physics, Kohn Hall, University of California, Santa Barbara, CA 93106;
bildsten@kitp.ucsb.edu}

\begin{abstract}
We present a new mode of hydrogen burning on neutron stars (NSs)
called diffusive nuclear burning (DNB).  In DNB, the burning occurs in
the exponentially suppressed tail of hydrogen that extends to the
hotter regions of the envelope where protons are readily captured.
Diffusive nuclear burning changes the compositional
structure of the envelope on timescales $\sim 10^{2-4} \,{\rm yrs}$,
much shorter than otherwise expected.  This mechanism is applicable to
the physics of young pulsars, millisecond radio pulsars (MSPs) and
quiescent low mass X-ray binaries (LMXBs).
\end{abstract}

\section{Introduction}

The composition of NS envelopes affects their cooling (Potekhin,
Chabrier \& Yakovlev 1997) and thermal radiation (Romani 1987). The
amount of material needed to change the spectral profile of the
thermal radiation from the surface of a NS is miniscule ($\sim
10^{-20} \Msun$).  Such a small amount of contamination could easily
be produced from spallation (Bildsten, Salpeter \& Wasserman 1992) of
fallback material (Woosley \& Weaver 1995) soon after a supernova
explosion.

Recent X-ray observations of the thermal spectrum from young radio
pulsars have yielded tantalizing clues of their photospheric makeup
(see Pavlov, Zavlin \& Sanwal 2002 for a review).  The thermal
emission from young NSs can be fit with two models: magnetic hydrogen
atmospheres or blackbody atmospheres.  Both atmospheres fit the
thermal spectra equally well, however they yield different effective
temperatures and different solid angles for the emission area.  The
model which yields a more reasonable radius for the NS at the
preferred distance is then taken as the favored model (Pavlov
et al. 2002).  For radio pulsars younger than $\sim 10^{4-5}\,{\rm
yrs}$ (e.g. Vela), the radius is reasonable for a magnetic hydrogen
atmosphere model (Pavlov et al. 2001).  For pulsars older than $\sim
10^{4-5}\,{\rm yrs}$ (e.g. PSR B0656+14), a blackbody model is favored
(see Pavlov et al. 2002 and references therein). This suggests a
possible evolution of hydrogen to more blackbody like elements on a
timescale of $10^{4-5}\,{\rm yrs}$, and these indications have
motivated our work.

At the photosphere, the local temperature ($T \sim 10^6\,{\rm K}$) and
density ($\rho \sim 1 \,{\rm g}\,{\rm cm}^{-3}$) are too low to allow
any significant nuclear evolution over $10^{4-5}\,{\rm yrs}$.  However
at a depth 1 m underneath the photosphere, the temperature is roughly
two orders of magnitude greater ($T \sim 10^8\,{\rm K}$).  Consider a
NS envelope of hydrogen on carbon (we use C here as our fiducial
proton capturing nucleus, results for other nuclei are in Chang \&
Bildsten 2003; CB03 hereafter) as shown in Figure 1.
\begin{figure}
\begin{center}
    \includegraphics[width=0.4\textwidth]{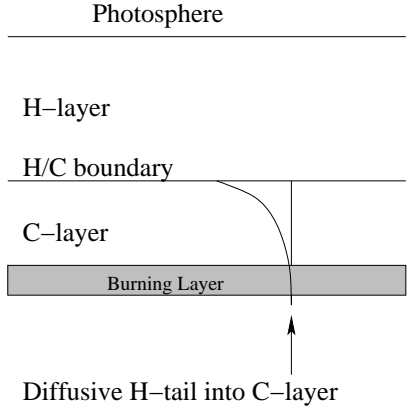}
    \caption{Schematic of a H/C envelope in diffusive equilibrium.
    The diffusive tail of hydrogen extends deep into the carbon,
    reaching temperatures where the hydrogen rapidly captures onto
    carbon, forcing depletion of the hydrogen layer as protons diffuse
    down to the burning layer.}
\end{center}
\label{fig:envelope_diagram}
\end{figure}
In diffusive equilibrium, the separation between the hydrogen and
carbon is not strict.  Rather a diffusive tail of hydrogen penetrates
deep into the carbon layer.  Protons easily reach a depth where the
temperature is sufficiently high for rapid capture onto C. Since the
hydrogen's diffusive tail is exponentially suppressed, there will be a
region where proton captures are peaked, which we call the burning
layer.  The consumption of the hydrogen by burning will drive the
diffusive tail out of equilibrium, which sets up a diffusive current
of hydrogen that flows down from the hydrogen layer into the burning
layer. We refer to this burning mechanism (first mentioned by Chui \&
Salpeter 1964 and initially calculated by Rosen 1968) as diffusive
nuclear burning (DNB; CB03).  Over time, given that there is no
interstellar accretion or other processes to refresh the hydrogen
layer, the H is depleted.

This scenario presents several competing timescales.  The first is the
proton capture timescale in the burning layer, $\tau_{\rm nuc}$.  The
second is the time it takes protons to diffuse into the burning layer,
$\tau_{\rm diff}$.  In the case where $\tau_{\rm diff} \ll \tau_{\rm
nuc}$, the diffusive tail is always in equilibrium.  In the opposite
limit, the diffusive tail is modified by the burning.  For simplicity
we will only discuss the first case here.

The equilibrium structure of the diffusive tail is calculated from
hydrostatic balance for each ion,  
\begin{eqnarray}\label{eq:hb}
\frac {dP_i} {dr} &=& -n_i \left( A_i m_p g - Z_i e E\right), \\
\frac {dP_e} {dr} &=& -n_e\left(m_e g + e E\right),
\end{eqnarray}
where $P_i$, $n_i$, $A_i$, $Z_i$ are the pressure, number density,
atomic number and charge of the $i$'th ion species and $E$ is the
upward pointing electric field found by demanding charge neutrality,
$n_e = \sum n_i Z_i$.  The thermal structure is determined from the
constant flux equation,
\begin{equation}
\frac {dT} {dr} = -\frac {3 \kappa \rho} {16 T^3}T_e^4, 
\label{eq:flux}
\end{equation}
where $\kappa$ is the opacity and $T_e$ is the effective temperature.
Given appropriate microphysics, the equilibrium thermal and
compositional structure can be calculated (see CB03 for additional
details).  The microphysics we have chosen in this paper are valid for
the non-magnetic case, $B < 10^9 \,{\rm G}$.

\begin{figure}
\begin{center}
    \includegraphics[width=0.55\textwidth]{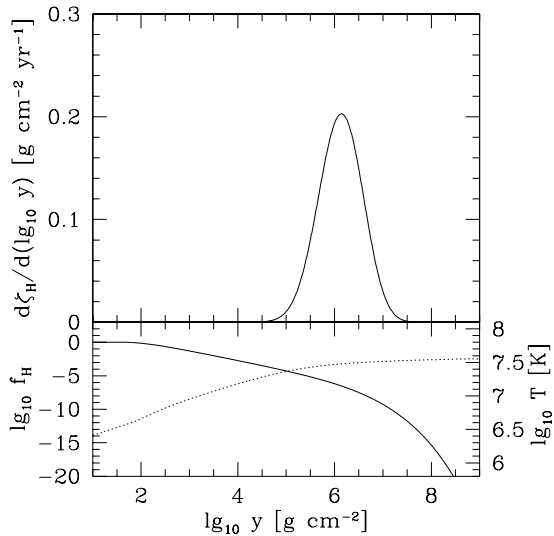}
  \caption{Differential hydrogen column burning rate taking into
    account p-p capture and p + C capture.  The bottom graph shows the
    number fraction (solid line) and temperature (dotted line).  This
    model has $y_H = 100 \,{\rm g\,cm}^{-2}$ and $T_e = 8 \times 10^5
    \,{\rm K}$.  The integrated burning rate for this model is
    $y_H/\tau_{\rm col} = 0.24 \,{\rm g\,cm}^{-2}\,{\rm yr}^{-1}$,
    giving $\tau_{\rm col} = 417 \,{\rm yrs}$.}
\end{center}
\end{figure}
The local hydrogen burning rate is $m_p n_H/\tau_{\rm nuc}$, hence the
total hydrogen burning rate per area on a NS, $\zeta_H$, is
\begin{equation}
\zeta_H = \frac {y_H} {\tau_{\rm col}} = 
\int \frac {n_H m_p} {\tau_H(n_H, n_C, T)} dz,
\label{eq:col_burning_rate}
\end{equation}
where $y_H = \int m_p n_H dz$ is the integrated column of hydrogen and
$\tau_{\rm col}$ is the characteristic burning time for that column.
In Figure 2, we show the equilibrium structure for a fiducial NS model
with $T_e = 8 \times 10^5\,{\rm K}$ and the total burning rate per log
column, $y = P/g$. The Gaussian peak in the burning rate traces out
the burning zone which is centered around a column of $y \approx
10^6\,{\rm g\,cm}^{-2}$, which is about 20 cm below where most of the
H resides.

\begin{figure}
\begin{center}
  \includegraphics[width=0.55\textwidth]{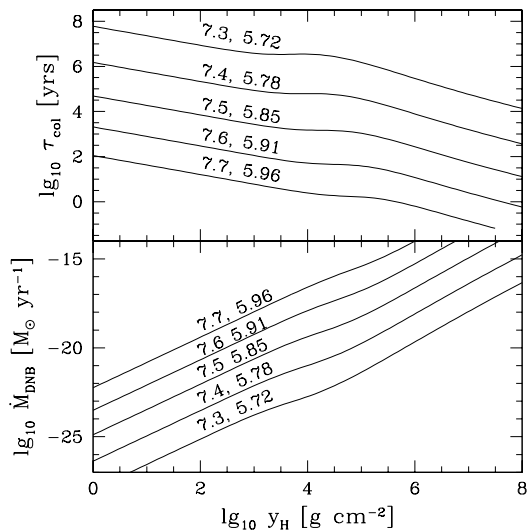}
    \caption{Characteristic burning time, $\tau_{\rm col}$, and total
      mass burning rate, $\dot{M}_{\rm DNB}$ for a NS radius of 10 km,
      as a function of $y_H$ for different fixed core temperatures and
      an underlying carbon layer.  For each model, we list the
      logarithmic core temperature and associated logarithmic
      effective temperature.}
\end{center}
\end{figure}
We relate the burning rate per column into a total mass burning rate
$\dot{M}_{\rm DNB} = 4\pi R_{*}^2 \zeta_H$.  In Figure 3, we plot the
characteristic burning time $\tau_{\rm col}$ and associated
$\dot{M}_{\rm DNB}$ as a function of the total hydrogen column, $y_H$.
For a given envelope with an initial column of hydrogen at a fixed
core temperature, the evolution follows the curve up to the
photosphere at $y_H \sim 1\,{\rm g\,cm}^{-2}$.  Because the evolution
follows a simple power law, the lifetime for any given envelope is
completely determined by the characteristic burning time, $\tau_{\rm
col}$, at the photosphere.  The curve for a central temperature of
$T_c = 5 \times 10^7\,{\rm K}$ is cut off at large columns since the
energy release from proton captures is comparable to the flux leaving
the envelope.  Hence our assumption of constant flux is violated.
What is also notable about Figure 3 is that the curves follow a power
law dependence which we derive in CB03.

Future directions for DNB are including the microphysics for magnetic
NSs, calculating DNB without the constraint $\tau_{\rm diff} \ll
\tau_{\rm nuc}$ and allowing for an intervening helium layer.  DNB
allows for young neutron stars to have atmospheres other than hydrogen
soon after birth.  For instance, NSs with carbon, nitrogen or oxygen
photospheres can easily exist after burning off any initial H.  DNB
may help explain the recent observation of magnetic oxygen lines on
1E1207+56 (see De Luca's review in this volume, also see Hailey \&
Mori 2002).

\acknowledgements

This research was supported by NASA via grant
NAG 5-8658 and by the NSF under Grants PHY99-07949 and
AST01-96422. L. B. is a Cottrell Scholar of the Research Corporation.

\end{document}